# MP3: A More Efficient Private Presence Protocol

Rahul Parhi Michael Schliep Nicholas Hopper
University of Minnesota
Minneapolis, MN
{parhi003,schli116,hoppernj}@umn.edu

## ABSTRACT

This paper proposes MP3, the second privacy-preserving presence protocol that leaks no information about the graph structure of the social network. Several cryptographic techniques are applied to improve the existing DP5 protocol—the first privacy-preserving presence protocol—while maintaining the same level of privacy. The key contribution of this paper is the use of a dynamic broadcast encryption scheme to reduce the size of the presence database. This enables cheaper registration and lookup required for the protocol. As compared to DP5, MP3 requires on the order of ten times less bandwidth of the servers during registration, and requires on the order of two times less bandwidth for lookup, for a small number of users ($N = 10000$). Furthermore, these savings asymptotically increase with the number of users. The client-side latency is also improved significantly in MP3, as compared with DP5. We provide an evaluation of the performance and scalability of both protocols.

## 1. INTRODUCTION

An often overlooked requirement of providing secure communication is protecting the metadata of the users. Due to the recent threat and concern of government surveillance and collection of user data [1], an increasing number of services have appeared with the goal of protecting users' privacy from the provider of the service. A common approach is end-to-end secure messaging, which is currently employed in services such as Apple's iMessage, Open Whisper Systems' Signal Protocol, and Facebook's Messenger [2–4]. Secure messaging hides the content of the conversation from the provider by using strong cryptographic techniques, but that is not enough as the metadata (who is contacting whom, when are they contacting each other, etc.) of the conversation is still known to the service provider.

A critical part of secure messaging is presence, i.e., knowing when a friend is online. Even though the secure messaging providers do not have access to the plaintext conversation, they still keep track of the set of friends for every user in order to run their service. This means that these services know the entire graph structure of their social network as well as the presence status of every user. To have a truly private communication platform, the communication metadata must also be protected.

To address this issue Borisov, Danezis, and Goldberg proposed DP5—the Daghstul Privacy Preserving Presence Protocol P—the first privacy-preserving protocol to leak no information about the social graph to third parties and limit the information retained by the service itself [5]. DP5 is a private presence protocol that allows the users to see the online status of their friends in a completely private manner. DP5 makes use of Private Information Retrieval (PIR) [6] for querying the service for a given user's buddy's presence.

The major challenge with DP5 is its scalability (or lack thereof). The presence database of DP5 grows much more rapidly than the number of users of the service. This creates a very expensive service for even a small number of users. To correct this issue, we propose MP3—the Minnesota Private Presence Protocol—as an improvement to DP5. MP3 significantly reduces the size of the presence database by leveraging a dynamic broadcast encryption scheme [7]. As a result, MP3 is significantly cheaper to run for even a relatively modest user base, with savings that increase as the number of users scales up.

**Key Contributions.** We improve the existing DP5 protocol by using a dynamic broadcast encryption scheme, while maintaining the same level of privacy. This enables a significant reduction in the size of the presence database, allowing much cheaper registration and lookup queries in the context of the bandwidth required from both a client and server-side perspective in the proposed MP3 protocol. The user-facing latency of MP3 is also an order of magnitude less than that of DP5.

This paper is organized as follows: Section 2 describes the background, goals, and related work pertaining to the MP3 protocol, Section 3 presents a detailed description of the MP3 protocol, and Section 4 analyzes the performance of MP3 and compares it to that of DP5.

## 2. BACKGROUND

The primary functionality of a private presence protocol is to allow for the registration of one's online presence and to query the presence status of one's buddies. Some auxiliary functionality is needed for MP3 and DP5 to achieve a completely private presence protocol. In our design, a user's buddies are a one way relationship, i.e., Alice follows Bob, but Bob does not necessarily need to follow Alice. Alice controls who may follow her by giving them explicit access (by sharing a key out-of-band) and has the ability to revoke that access at any time.

### 2.1 DP5 Overview

Since our design shares many characteristics with DP5, we give a brief overview of the protocol here. DP5 makes use of Private Information Retrieval (PIR), in which a database is distributed to several servers so that a user can query the servers to retrieve a specific record without revealing which record they retrieve. Given this functionality, a "trivial" private presence protocol would have each user $A$ with $n_A$ friends encrypt $n_A$ *presence records* recording their status (and possibly other information, such as a contact address), with the shared key for each friend, and periodically upload this information to a presence database. When $A$'s friend $B$ wants to check on $A$'s status, they would query the current database (using PIR) for the presence records encrypted under the symmetric key $A$ shares with $B$. To hide information about the social graph, each user would need to pad the number of presence records uploaded per period to some maximum value denoted as $N_{\text{fmax}}$. Since the server computational costs of PIR scale linearly in the size of the database (and the bandwidth costs also increase, though sub-linearly) this protocol would result in a quadratic-size database.

To combat this problem, in DP5 the presence service is split into two asymmetric services. The primary, short-term service is used to register and query presence of users with the precision of short (on the order of 5 minute) windows. The secondary service is the long-term database, which is used to provide metadata for querying during the short-term. The long-term database also provides friend registration and revocation with the precision of long (on the order of 24 hour) windows. As in the "trivial" protocol, it is assumed that users share a unique secret with each of their friends (most probably through a Diffie-Hellman key exchange [8]). In order to not leak information about how many friends a given user has DP5 (and MP3) defines the limit of $N_{\text{fmax}}$ on the number of friends a user may have.

In each period of the long-term database, a user (let's say Alice) of DP5 will upload her presence to the registration mechanism of DP5. This is referred to as Alice's long-term presence record. This record is actually *several* records, one for each of Alice's friends, padded to the limit on the number of friends a user may have. Usually, this limit will be on the order of the number of users of the service, meaning the long-term presence database scales almost with the square of the number of users, which in turn increases the amount of bandwidth and CPU this service requires from both client and server-side perspectives. These long-term presence records contain an ephemeral symmetric encryption key and signature verfication key that Alice then uses to encrypt and sign a *single* presence record for any short-term period in which she is online, which is uploaded to the short-term presence database. Thus the short-term database, which is queried more often, grows only linearly with the number of online users.

We address the issue of scaling in the long-term database by reducing the number of records Alice uploads in each long-term database to just a single record, similar to the short-term database, by using a dynamic broadcast encryption scheme.

## 2.2 Threat Model

We make standard assumptions about the users and adversaries of MP3. They are real world adversaries with common capabilities.

- We assume that honest users' systems are secure and not compromised. We also assume that honest servers can maintain secrecy and integrity. Our design maintains forward secrecy and does not require servers to store any long-term secrets.
- We allow the adversary to be an observer or a dishonest user of the system, and we assume they have not made any recent breakthrough in computational cryptographic assumptions, and assume that they cannot distinguish between different ciphertexts (IND-CPA). More detailed cryptographic assumptions are described in Section 3.
- Our security properties are under the covert model, i.e., adversaries will not act dishonestly if it would cause them to be detected and identified.
- Our protocol uses a robust information-theoretic PIR (IT-PIR) scheme [9] that requires multiple servers. This scheme requires that at least one server does not collude. This is described in more detail in Section 3.1.3.

## 2.3 Security Goals

The primary goal of MP3 is to provide presence information in a private manner. Here we describe the goals required for a private presence service.

**Privacy of presence and auxiliary data.** The presence status of a user and their auxiliary data should be available to only the explicit friends of that user.

**Integrity of presence and auxiliary data.** The friends of Alice should not accept the presence and auxiliary data unless it was submitted by Alice.

**Unlinkability of sessions.** A user should not be able to be linked between uses of MP3. The infrastructure and non-friend users should not be able to link the presence of another user between epochs.

**Privacy of the social graph.** No information about the social graph of a user should be revealed to any other party of MP3. More specifically, friends should not learn about other friends and the infrastructure should not learn any new information about a user.

**Forward/Backward secure.** Any compromised keys should not reveal past or future information that is secured with previous or future keys.

**Auditable** All operations performed by the infrastructure should be auditable. A user should detect when their friend registration or presence registration has not been performed honestly by the service provider.

**Support for anonymous channels.** The protocol should not require any identifying information for operation. The use of an anonymous channel should only enhance the privacy of the system, and MP3 will not compromise the anonymity of the user.

Lastly, DP5 claims indistinguishably of offline status and revocation. We show in Section 5.2 that there are severe limitations on the extent to which a presence protocol can conceal revocation status. As a result, we do not try to maintain this goal, which allows us our approach.

## 2.4 Related Work

Excluding DP5, there does not exist any previous work addressing the privacy of presence and social graphs. Dissent [10] and Riposte [11] offer anonymous micro-blogging services; these systems are similar to private presence in that posting a micro-blog implies the author was online. Dissent is based on a DC-net with a client-server architecture. Clients in Dissent must form a group to post anonymous messages for each other using distrusted servers. Dissent provides anonymity within a static group. Riposte utilizes a novel private database writing mechanism based on techniques of PIR. Both of these systems have high latency when dealing with large anonymity sets and are not concerned with the privacy of the social graph.

# 3. THE MP3 PROTOCOL

At a high level, the purpose of MP3 is to provide an efficient presence mechanism that leaks no information to any third-party about the graph structure of the social network (i.e., who is friends with whom). At this level, a user (call her Alice) will register her presence with the *registration mechanism* of MP3; one of her friends (call him Bob) can then query for Alice's presence with the *lookup mechanism* of MP3, and retrieve the status of Alice's presence. By employing several cryptographic techniques, this process can be carried out without leaking any information.

For the remainder of this paper, denote Alice (labeled $a$) as a user that is registering her presence, and Bob (labeled $b$) as a user that is querying for Alice's presence.

## 3.1 Overview

To register her presence, Alice must encrypt a *presence record* and register it through the registration mechanism of MP3. This record must be encrypted in such a manner that only the set of Alice's friends can decrypt it. For Bob to query, or lookup, the presence of Alice, he must request the record from the lookup mechanism of MP3 that was uploaded by Alice. If this lookup was done naïvely, the lookup mechanism would learn that Bob is requesting Alice's presence record. To circumvent this, the lookups are done using Private Information Retrieval (PIR). The registration mechanism of MP3 uses a single registration server, and the lookup mechanism is a set of $N_{\text{lookup}}$ PIR lookup servers.

The use of PIR requires time to be divided into discrete units, referred to as *epochs*. At the end of an epoch, the registration mechanism of MP3 compiles all presence records registered during that epoch and creates a PIR database. This database (and the metadata associated with it) is then sent to each of the $N_{\text{lookup}}$ lookup servers. The specifics of PIR are detailed in Section 3.1.3.

In order for Bob to know which record to request from the lookup mechanism, Alice and Bob must have a shared identifier that is included in Alice's presence record. In the event that an adversary learns this identifier, they can now query for Alice's presence indefinitely. To circumvent this, *long-term* and *short-term* epochs are employed, and by congruence, different sets of a single registration server and $N_{\text{lookup}}$ PIR lookup servers are employed for long-term epochs and short-term epochs. Ephemeral *epoch keys* are also employed such that new epoch keys are generated every long-term epoch. Thus, in the event of a compromise of these keys, only the presences for *that* epoch are compromised, and future epochs are safe as new keys are generated in the next long-term epoch.

### 3.1.1 Primitives

It is assumed that everyone participating in MP3 shares a set of known cryptographic primitives. Let $G_1$ and $G_2$ be two additive (written multiplicatively) cyclic groups of prime order $p$ and let $G_T$ be a multiplicative cyclic group also of prime order $p$. Denote $\mathbb{Z}_p$ as the ring of integers modulo $p$. An efficiently computable asymmetric pairing [12] defined by the map $e : G_1 \times G_2 \to G_T$ is known such that for generators $g_1 \in G_1$, $g_2 \in G_2$, and $\forall u, v \in \mathbb{Z}_p$:

$$e(g_1^u, g_2^v) = e(g_1, g_2)^{uv} \in G_T$$

The discrete logarithm problem (DLP), computational Diffie-Hellman problem (CDH), and the computational co-Diffie-Hellman problem (co-DHP or co-CDH) are assumed hard for $G_1$ and $G_2$. It is also assumed that we have a type 3 pairing so that efficiently computable isomorphisms from $G_1 \to G_2$ and $G_2 \to G_1$ do not exist [13].

Let $\mathcal{T}$ be the set of valid long-term epoch timestamps and $\tau$ be the set of valid short-term epoch timestamps. MP3 also makes use of the following:

- $\text{PRF}_K : \tau \to \{0,1\}^\alpha$, a keyed pseudorandom function where $K \in \{0,1\}^\beta$ that maps a short-term epoch timestamp to bit-strings of length $\alpha$.
- $H_0 : \tau \to G_2$, an efficiently computable hash function that maps short-term epoch timestamps to elements of $G_2$.
- $H_1 : G_1 \to \{0,1\}^\beta$, an efficiently computable hash function that maps elements of $G_1$ to bit-strings of length $\beta$; that is, keys in the pseudorandom function.
- $H_2 : \{0,1\}^\delta \to \{0,1\}^\ell$, an efficiently computable hash function that maps bit-strings of length $\delta$ to bit-strings of length $\ell$. $\delta$ is the length of a long-term public key and $\ell$ is the length of an identifier.
- $H_3 : G_T \to \{0,1\}^\ell$, an efficiently computable hash function that maps elements of $G_T$ to bit-strings of length $\ell$. Here again, $\ell$ is the length of an identifier.
- $\text{AEAD}_K^{\text{IV}}(h, m)$, an authenticated encryption function where $K \in \{0,1\}^\alpha$ (not to be mistaken for the key in the PRF), $m$ is the message, and $h$ is an unencrypted header used to derive a message authentication code (MAC). When $h = \varepsilon$, the empty string, it will be left out.

### 3.1.2 Epoch Specifics

Long-term epochs are meant to last on the order of a day, while short-term epochs are meant to last on the order of several minutes. The purpose of the long-term epoch is for a user to query the long-term lookup servers for their buddies' ephemeral epoch keys. The purpose of the short-term epoch is to enable a user to query for their buddies' presence. The ephemeral epoch keys are a tuple containing a buddy's shared identifier for the *next* long-term epoch and the shared identifier for the set of current[1] short-term epochs. Define the following notation:

- $T_j \in \mathcal{T}$, the $j$'th long-term epoch.
- $t_i \in \tau$, the $i$'th short-term epoch.
- $P_a^j$, the long-term identifier for the $j$'th long-term epoch for buddy Alice; call this Alice's long-term public key.
- $p_a^j$, the short-term identifier for all short-term epochs *during the j'th long-term epoch*; call this Alice's public presence key.

[1]current as in all short-term epochs during *this* long-term epoch

If Bob were to query the lookup servers during $T_j$ for his buddy Alice (using Alice's long-term public key, $P_a^j$, that he previously queried during epoch $T_{j-1}$), he would receive Alice's encrypted epoch keys and decrypt them to get $(P_a^{j+1}, p_a^{j+1})$. Because of this setup, Bob (and the rest of Alice's buddies) will learn both her long-term public key and public presence key for a given epoch. This means that there is nothing stopping one of Alice's buddies from uploading a record as her. To prevent this, we employ two signature schemes, one for long-term presence records and another for short-term presence records. Thus, all of Alice's friends can be confident that the record they received from MP3's lookup mechanism can only belong to Alice. During long-term epochs, we employ an elliptic curve digital signature [14], specifically using the Edwards-curve Digital Signature Algorithm (EdDSA), we make use of the reference implementation, Ed25519 [15]. During short-term epochs, the Boneh-Lynn-Shacham (BLS) [16] signature scheme is employed.

Long-term epochs are essentially a broadcast encryption scheme [17]. Briefly, broadcast encryption is a mechanism in which a *broadcast manager* distributes some sort of encrypted message to a set of known users who are qualified to decrypt the message, while maintaining the ability to revoke and add other users to this set at any time. In the context of MP3's long-term epoch, every user is a broadcast manager, their encrypted message is their epoch keys, and the set of qualified users is their friends. We implement the long-term epoch using a *dynamic broadcast encryption* scheme with constant-size ciphertexts and decryption keys [7].

Short-term epochs are solely for getting your buddies' current presence. The precision in this is determined by the length of the short-term epoch. Just the existence of a short-term record for your buddies implies that they are online, but they may wish to include a presence message including information such as their encryption public key, how to make contact with them, etc. Denote $m_a^i$ as the presence message of Alice for epoch $t_i$.

### 3.1.3 Private Information Retrieval

In order to not reveal which presence records are being retrieved when a user performs a lookup, PIR is used. Specifically, *information-theoretic* PIR (IT-PIR), which requires multiple ($N_{\text{lookup}}$) PIR lookup servers. IT-PIR is used over *computational* PIR (C-PIR) due to its large speed improvement. In the event C-PIR was used, only a single lookup server would be required. In IT-PIR, a user splits their query across each of $N_{\text{lookup}}$ lookup servers, and combines the responses from all the servers to retrieve the requested data. We implement the same hash based PIR protocol [6,9,18,19] as in DP5, where the database is a hash table storing a map of ⟨key,value⟩ pairs. For example, when querying for Alice, the key is an identifier of length $\ell$ and the value is the Alice's presence record. If we have $n$ presence records, each of length $s$ (in bytes) at an epoch change, we create $r = \lceil \sqrt{ns} \rceil$ buckets. The expected value of the size of each bucket is $\frac{n}{r}$ records and probabilistically no more than $\frac{n}{r} + \sqrt{\frac{n}{r}}$.

To determine the hash function used to hash the keys for the database in such a way that the size of the database is minimized, ten random keys for a PRF $\Pi^{(r)}$ with image $\{1, \ldots, r\}$ are selected, using each to hash all the presence records. The key that yields the smallest largest bucket is used to define the hash function. All buckets are padded to the size of this smallest largest bucket. This operation is done by the registration server at the end of every epoch, after which, the resulting databases are then sent off to each lookup server.

## 3.2 Setup

To participate in MP3, Alice randomly selects $G_a \in G_1$, $H_a \in G_2$, and $\gamma_a \in \mathbb{Z}_p$. She then stores the tuple $(G_a, H_a, \gamma_a)$ as her *manager key*.

For every friend $f$ of Alice, she randomly selects $x_{af} \in \mathbb{Z}_p$ and stores it. She then derives the following from her manager key:

$$A_{af} = G_a^{\frac{x_{af}}{\gamma_a + x_{af}}} \qquad B_{af} = H_a^{\frac{1}{\gamma_a + x_{af}}}$$

The tuple $(x_{af}, A_{af}, B_{af})$ is then shared out-of-band with friend $f$ as their *decryption key* (associated with Alice) along with her current long-term public key, $P_a^j$.

## 3.3 Long-term Epoch

### 3.3.1 Registration

To register for epoch $T_j$, Alice must register during epoch $T_{j-1}$. To begin, Alice generates a new long-term public-private key pair as $(P_a^j, Y_a^j)$ where $Y_a^j$ is a randomly selected private key and $P_a^j$ is its associated point on Ed25519. She also generates a new public-private presence key pair as $(p_a^j, y_a^j)$, where $y_a^j \in \mathbb{Z}_p$ and $p_a^j = g_1^{y_a^j}$, $g_1 \in G_1$.

During a long-term epoch, Alice has the opportunity to revoke a certain number of her buddies. Denote this as $N_{\text{rev}}$ and note that it is a parameter of the MP3 protocol and constant for all users. Using the stored $x$ values from the decryption keys of the set of buddies Alice wishes to revoke and her manager key, she computes $R_a^j$ and updates her manager key as follows:

**Algorithm 1** Compute $R_a^j$ and update Manager Key

```
1: R = {x_r | r = 0, 1, ..., N_rev}  // x values to revoke
2: R_a^j := ε
3: for each x ∈ R do
4:     B := H_a^(1/(γ_a + x))
5:     R_a^j := R_a^j || x || B
6:     H_a := B
7: end for
```

Revocations are permanent, meaning if Alice would like to "re-friend" a previously revoked buddy, she would have to share a new decryption key with them using the construction in Section 3.2. In the event that Alice would like to revoke less than $N_{\text{rev}}$ buddies, she can just pad $R_a^j$ with random buddies up to $N_{\text{rev}}$.

Alice then selects a random $\kappa \in \mathbb{Z}_p$ and computes the following:

$$C_{a1}^j = G_a^{\kappa\gamma_a} \ , \qquad C_{a2}^j = H_a^{\kappa} \ , \qquad K_a^j = e(G_a, H_a)^{\kappa}$$

She then encrypts her epoch keys as:

$$C_a^j = \text{AEAD}^j_{K_a^j}(P_a^{j-1} \parallel R_a^j \parallel C_{a1}^j \parallel C_{a2}^j, P_a^j \parallel p_a^j)$$

Finally, she creates the signature:

$$S_a^j = \text{sign}_{Y_a^{j-1}}(P_a^{j-1} \parallel R_a^j \parallel C_{a1}^j \parallel C_{a2}^j \parallel C_a^j)$$

Alice then uploads $P_a^{j-1} \parallel R_a^j \parallel C_{a1}^j \parallel C_{a2}^j \parallel C_a^j \parallel S_a^j$ to the long-term registration server.

Upon receiving Alice's record, the long-term registration server will first verify $S_a^j$ with the record using $P_a^{j-1}$. If the signature is valid, the server will compute $ID_a^j = H_2(P_a^{j-1})$, and store $\langle ID_a^j, R_a^j \parallel C_{a1}^j \parallel C_{a2}^j \parallel C_a^j \parallel S_a^j \rangle$.

### 3.3.2 Lookup

To lookup Alice's presence for epoch $T_j$, Bob first requests the metadata associated with the databases from each lookup server. Since all lookup servers have the same database, the metadata should be the same, but in the event that some of the servers are dishonest, he takes the majority of the received metadata. The metadata contains information about the number of buckets and size of the buckets. He then computes $ID_a^j = H_2(P_a^{j-1})$, and builds a PIR request using the metadata for $ID_a^j$ (and the rest of his buddies), splits his query across the $N_{\text{lookup}}$ long-term lookup servers, combines the responses[2] from each server and retrieves:

$$R_a^j \parallel C_{a1}^j \parallel C_{a2}^j \parallel C_a^j \parallel S_a^j$$

Bob then verifies the signature with $P_a^{j-1}$ (to be absolutely certain that the record is from Alice). If the signature is invalid, he does not complete the steps outlined in the remainder of this section.

Bob updates his decryption key as follows:

**Algorithm 2** Update Decryption Key

1: $\mathcal{R}' = \{(x_r, B_r) \mid r = 0, 1, \ldots, N_{\text{rev}}\}$ // from $R_a^j$
2: **for each** $(x, B) \in \mathcal{R}'$ **do**
3: $\quad B_{ab} := \left(\frac{B}{B_{ab}}\right)^{\frac{1}{x_{ab}-x}}$
4: **end for**

In the event that Bob is in $\mathcal{R}'$, he cannot do the computation on line 3 as he would try to compute 1/0. If he chose to not update his $B_{ab}$, he would not be able to properly decrypt $C_a^j$, thus he is permanently revoked from Alice's buddy list.

Assuming Bob has not been revoked, he then computes[3]:

$$\begin{aligned} & e(C_{a1}^j, B_{ab}) \cdot e(A_{ab}, C_{a2}^j) \\ &= e(G_a^{\kappa\gamma_a}, H_a^{\frac{1}{\gamma_a + x_{ab}}}) \cdot e(G^{\frac{x_{ab}}{\gamma_a + x_{ab}}}, H_a^{\kappa}) \\ &= e(G_a, H_a)^{\frac{\kappa\gamma_a}{\gamma_a + x_{ab}}} \cdot e(G_a, H_a)^{\frac{x_b\kappa}{\gamma_a + x_{ab}}} \\ &= e(G_a, H_a)^{\frac{\kappa\gamma_a + \kappa x_{ab}}{\gamma_a + x_{ab}}} \\ &= e(G_a, H_a)^{\kappa} = K_a^j \end{aligned}$$

and finally uses $K_a^j$ to decrypt $C_a^j$ retrieving $P_a^j$ and $p_a^j$.

To prevent the lookup mechanism of MP3 from learning how many buddies Bob has, he must make a lookup for $N_{\text{fmax}}$ buddies. Note that $N_{\text{fmax}}$ is another parameter of the protocol and is constant for all users. If Bob has less than $N_{\text{fmax}}$ buddies, he pads his request with random IDs up to $N_{\text{fmax}}$.

[2]the PIR protocol we use [9] supports detection of malicious servers. The presence data is reconstructable when only $t+2$ servers are honest, and $t$ is the privacy level. Any collusion of upto $t$ servers learns no information about the presence data.

[3]This computation has been simplified to the case in which Alice has not revoked anyone and thus not updated her manager key and Bob has not updated his decryption key. In the case that the keys have been updated, the exponents of the group elements will be the same, but the group elements themselves will be different, so this computation is equivalent.

## 3.4 Short-term Epoch

### 3.4.1 Registration

To register for epoch $t_i$, Alice must register during $t_{i-1}$. Assume that epoch $t_i$ is during epoch $T_j$. To begin, Alice encrypts her presence message, $m_a^i$ as follows:

$$k_a^i = \text{PRF}_{H_1(p_a^j)}(t_i) \qquad c_a^i = \text{AEAD}_{k_a^i}^i(m_a^i)$$

She then computes the unforgeable signature:

$$s_a^i = H_0(t_i)^{y_a^j}$$

Alice then uploads $c_a^i \parallel s_a^i$ to the short-term registration server.

Upon receiving Alice's record, the short-term registration server will compute $id_a^i = H_3(e(g_1, s_a^i))$, and store $\langle id_a^i, c_a^i \rangle$.

### 3.4.2 Lookup

To lookup Alice's presence for epoch $t_i$, Bob requests the metadata associated with the short-term databases in the same manner as in the long-term epoch, then computes $id_a^i = H_3(e(p_a^j, H_0(t_i))$ which is equivalent to the registration server's computation of $H_3(e(g_1, s_a^i))$ by the properties of pairings. He then builds a PIR request for $id_a^i$ (and the rest of his buddies), splits his query across $N_{\text{lookup}}$ short-term lookup servers, combines the responses from each server and retrieves $c_a^i$. Bob can be absolutely certain that $c_a^i$ is from Alice due to the unforgeable signature $s_a^i$.

Bob can then compute $k_a^i = \text{PRF}_{H_1(p_a^j)}(t_i)$ and decrypt $c_a^i$ retrieving $m_a^i$. The decryption being successful implies that Alice is online during epoch $t_i$. As in the long-term epoch, in order to not leak information of how many buddies Bob has, he must pad his lookup to $N_{\text{fmax}}$ ids.

## 3.5 Remarks

### 3.5.1 Details

Recall the parameters $N_{\text{lookup}}$, $N_{\text{fmax}}$, and $N_{\text{rev}}$. Denote $N$ as the number of users. Long-term epochs should last on the order of a day and short-term epochs on the order of several minutes. The length of a short-term epoch defines the precision in querying for a given user's buddies' current presence. The length of a long-term epoch defines how often buddies can be revoked. The performance of the operations of both epochs is comparable.

The client-side space complexity for long-term registration is $\Theta(N_{\text{rev}})$ and for short-term registrations is $\Theta(1)$. Both the long-term and short-term databases store $N$ records. As described in Section 3.1.3, the databases are stored as a map of $\langle$key,value$\rangle$ pairs. Recall that we create $r = \lceil \sqrt{ns}\, \rceil$ buckets where $n$ is the number of records and $s$ is the length of a record. Each long-term presence record is inherently larger than a short-term presence record. Also recall that the upper bound of bucket size is probabilistically no more than $\frac{n}{r} + \sqrt{\frac{n}{r}}$. Because of the asymmetry in record lengths between short-term and long-term epochs, the number of buckets will be greater and the number of records per bucket will be less in a long-term database, than that of the short-term database. This makes the lookup cost in the long-term database slightly more than that of the short-term.

### 3.5.2 *Improvements to DP5*

In DP5, the server-side space complexity for long-term registration is $\Theta(N \cdot N_{\text{fmax}})$, because long-term database stores $N \cdot N_{\text{fmax}}$ records. This is due to Alice needing to upload a single presence record for each of her buddies. We improve the long-term registration by using a dynamic broadcast encryption scheme to require Alice to only upload a single presence record that all of her buddies query, reducing the space complexity to $\Theta(N \cdot N_{\text{rev}})$. We leave the short-term epoch of DP5 unchanged as it is already inexpensive. Our improvement greatly reduces the size of the long-term database by removing the scaling with $N_{\text{fmax}}$ for long-term presence records, and as such reduces the required bandwidth to run the service for both registration and lookup. Note that DP5 bandwidth costs do not rely on $N_{\text{rev}}$, but that is a non-issue, as argued in Section 5.1.

### 3.5.3 *Skipping Epochs*

In order to maintain privacy and not leak or expose any information, users have to follow a strict set of guidelines regarding epochs. It may often be the case that a user misses certain epochs. Due to short-term epochs being solely for retrieving the presence status for a users' buddies, it will often be the case that many users do not participate in every short-term epoch. Users can choose how often they wish to participate in short-term epochs based on non-sensitive information (such as network availability), but cannot be based on information from or related to previous epochs. Otherwise, an adversary could create a timing attack and learn information about a given user's buddies.

In the case of long-term epochs, skipping epochs creates a problem in that short-term queries rely on information from the previous long-term epoch. MP3 assumes that all users have queried all long-term databases in order to have the most recent long-term public keys for their buddies. To mitigate this, a user must query all previous long-term databases in order to retrieve the most recent long-term public keys for their buddies. A user cannot just query long-term databases until they find the buddy in question as the lookup mechanism or an observer would then learn that the last time *this* buddy was online was the last epoch the user queried. This process must be carried out for all long-term lookups, not just when one of your buddies misses an epoch as to not leak any information.

As this operation becomes very expensive as the length of time an MP3 service is running, we place a bound on how many long-term databases are stored (on the order of 30 days). This requires all users to participate in the long-term epoch at least once within this bound, or in the case that this is impossible, share their most recent long-term public key with all their buddies out-of-band to get back on MP3.

# 4. PERFORMANCE

## 4.1 Implementation

The MP3 protocol is built on top of HTTP. The core cryptography relies on OpenSSL for AES [20] and SHA-256 [21], RELIC [22] for pairing-friendly curves, Percy++ [23] for PIR, and Ed25519 for non-group signatures. The server-side networking relies on the Twisted network programming framework.

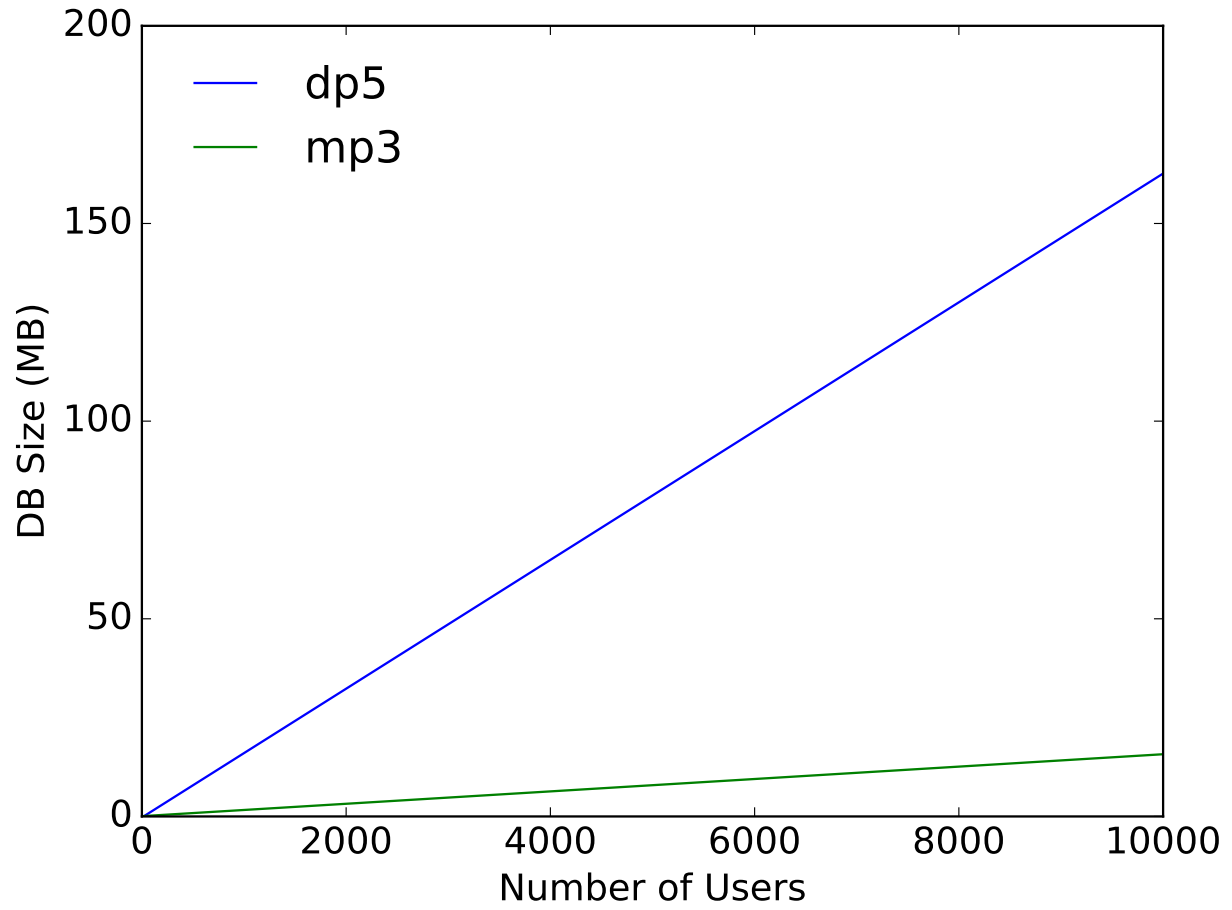

**(a) Varying $N$, holding $N_{\text{fmax}} = 100$ fixed.**

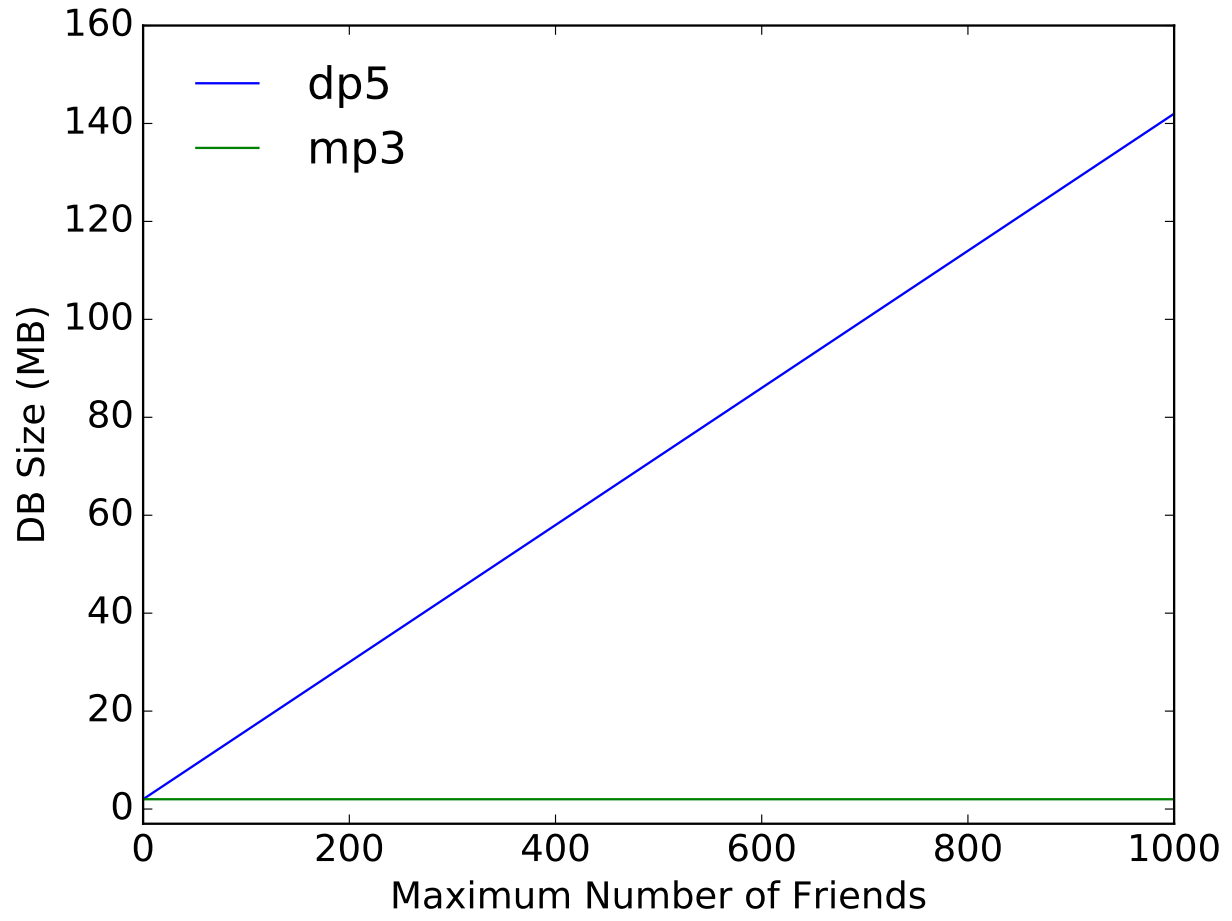

**(b) Varying $N_{\text{fmax}}$, holding $N = 1000$ fixed.**

**Figure 1: Long-term databases sizes for MP3 and DP5, $N_{\text{rev}} = 1$**

The groups $G_1$, $G_2$, and $G_T$ are defined by the Optimal Ate pairing over a 256-bit Barreto-Naehrig curve. We use Andrew Moon's implementation[4] of Ed25519 for deriving long-term public/private keys and long-term epoch signatures. AEAD is implemented using AES in Galois/Counter Mode (GCM) [24]. The PRF and hash functions are implmented using SHA-256. All client-side network interactions are encoded as HTTP requests using the `grequests`[5] module. The server-side networking is built using the Twisted Web framework. In all, our MP3 library is implemented in 4500 lines of C++ and 1400 lines of Python 2.7.

## 4.2 Experiments

To evaluate the performance of MP3 vs. DP5, we simulated both protocols in an event-driven manner with the number of users ranging from 100 to 2000 clients. To simulate the worst-case scenario, we had all clients perform registration and lookup for all epochs. We staggered when clients began an operation in order to avoid the thundering

[4]https://github.com/floodyberry/ed25519-donna
[5]https://github.com/kennethreitz/grequests

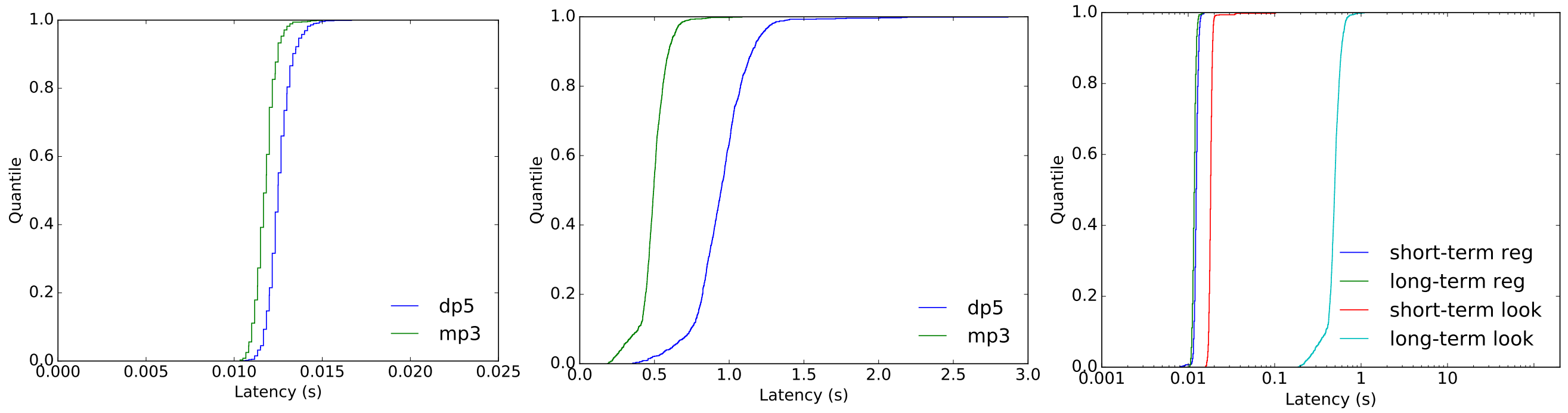

**(a) Long-term registration latency, MP3 vs. DP5.**

**(b) Long-term lookup latency, MP3 vs. DP5.**

**(c) Overall latency in MP3 for all client-side operations.**

**Figure 2: Overall user-facing latency comparing MP3 and DP5, including RTT.**

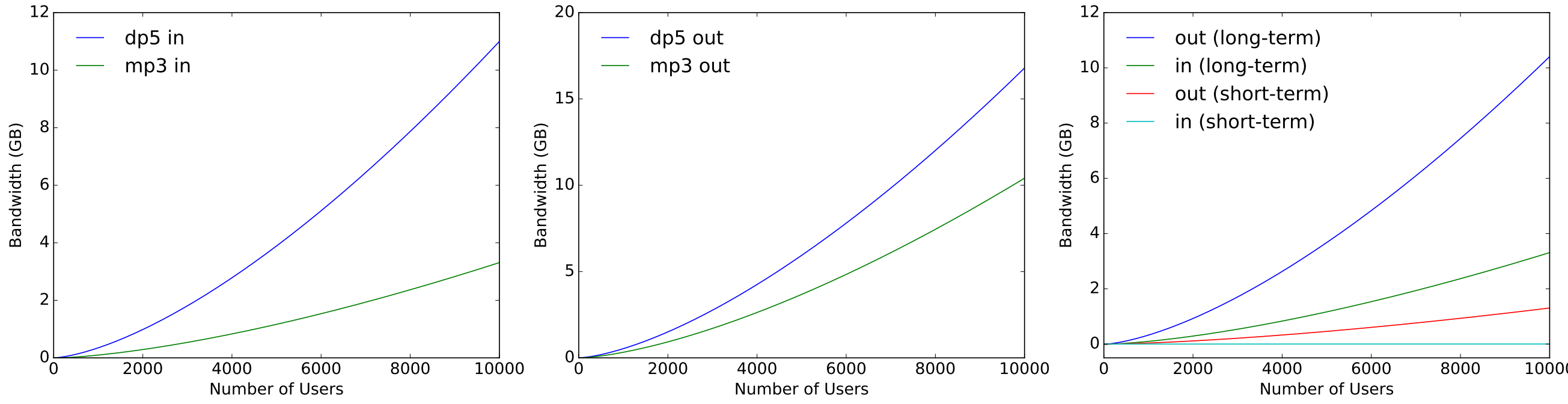

**(a) Inward bandwidth of a long-term lookup server, MP3 vs. DP5.**

**(b) Outward bandwidth of a long-term lookup server, MP3 vs. DP5.**

**(c) Overall bandwidth of long-term and short-term MP3 lookup servers.**

**Figure 3: All bandwidths for lookup servers, comparing MP3 and DP5. All data is for a single epoch. $N_{\mathbf{fmax}} = 100$, $N_{\mathbf{rev}} = 1$.**

herd problem. All simulations were run across several 3.4 GHz Haswell i7 machines, all with 32 GB of RAM. In each setup, clients were evenly spread across 8 machines, a single machine was used for both short-term and long-term registration, and each short-term and long-term lookup server received its own machine. $N_{\text{lookup}}$ was held fixed at 3 for all setups and both protocols. The equivalent of 6 months or 180 days (with 24-hour long-term epochs and 5 minute short-term epochs) of execution were simulated in all setups. For both MP3 and DP5, the most expensive components are the PIR lookup servers, both from a CPU and bandwidth point of view.

The main bottleneck with DP5 is the long-term lookup server due to the size of the database scaling with $N_{\text{fmax}}$. Figure 1 compares the long-term database sizes for MP3 and DP5, when the number of users is varied and $N_{\text{fmax}}$ is fixed, and when $N_{\text{fmax}}$ is varied and the number of users is fixed. Note that the database size for DP5 is inherently larger in both plots, and specifically grows rapidly when varying $N_{\text{fmax}}$, while the database size for MP3 is constant.

The user-facing latency for both protocols is summarized in Figure 2. Both protocols were simulated with 1000 users and $N_{\text{fmax}} = 100$; for MP3, $N_{\text{rev}} = 1$. Increasing $N_{\text{rev}}$ for MP3 will have no effect on the latency, so this comparison is valid. As seen in Figure 2a, the long-term registration latency between the two protocols is comparable as the operation for storing presence records doesn't change with the number of records uploaded by a user. As seen in Figre 2b, the long-term lookup latency is an order of magnitude higher for DP5. This is due to less records being processed by the lookup servers in MP3. The number of records in a DP5 long-term database scales with $\Theta(N \cdot N_{\text{fmax}})$, while the number of records in an MP3 long-term database only scales with $\Theta(N)$ for MP3. The short-term registration and lookup for both protocols were nearly identical, as expected.

**Table 1: Client bandwidth costs for a single epoch and comparison to common websites. $N = 1000$, $N_{\mathbf{fmax}} = 100$, $N_{\mathbf{rev}} = 1$.**

| | | MP3 | DP5 | twitter.com | amazon.com |
|---|---|---|---|---|---|
| IN | short-term | 49 KB | 48 KB | 3.3 MB | 4.3 MB |
| | long-term | 720 KB | 1122 KB | | |
| OUT | short-term | 184 B | 184 B | | |
| | long-term | 160 KB | 680 KB | | |

The overall client-side bandwidth is summarized in Table 1. The total bandwidth for a long-term epoch is significantly less for MP3 (880 KB) than that of DP5 (1802 KB) and compared with commonly visited websites such as twitter.com (3.3 MB) and amazon.com (4.3 MB), the bandwidth costs are very reasonable. As expected, the bandwidth costs in the short-term are nearly identical for both protocols, and much

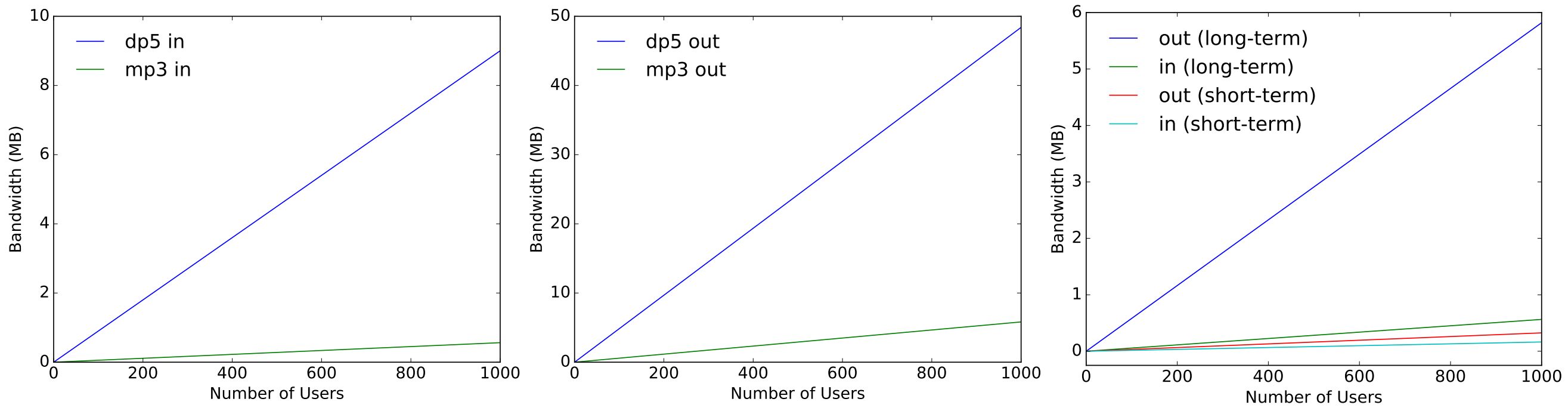

**(a) Inward bandwidth of a long-term registration server, MP3 vs. DP5.**

**(b) Outward bandwidth of a long-term registration server, MP3 vs. DP5.**

**(c) All bandwidths for long-term and short-term MP3 registration servers.**

**Figure 4: Overall bandwidth for registration servers, comparing MP3 and DP5. All data is for a single epoch. $N_{\mathbf{fmax}} = 100$, $N_{\mathbf{rev}} = 1$.**

smaller (49.2 KB) than either of twitter.com or amazon.com. Between both protocols, the client-side CPU usage was comparable, and scales linearly with the number of records in the database.

The bandwidth usage for the PIR lookup servers is summarized in Figure 3. Server-side lookup communication costs scale with the size of the database and the number of users, specifically, in the long-term database, $\Theta(N\sqrt{N \cdot N_{\text{fmax}}})$ for DP5 and $\Theta(N\sqrt{N \cdot N_{\text{rev}}})$ for MP3. Due to DP5's scaling relying on $N_{\text{fmax}}$, the bandwidth required to run an MP3 long-term lookup server is significantly less. This leads to improvements on the order of 2 times less bandwidth for a long-term lookup server with MP3 with $N = 10000$ users, $N_{\text{fmax}} = 100$, and $N_{\text{rev}} = 1$, and this disparity grows even more as both $N$ and $N_{\text{fmax}}$ increase. In the short-term, the communication costs scale as $\Theta(N\sqrt{N})$ for both protocols.

The bandwidth usage for the registration servers is summarized in Figure 4. The communication costs in the long-term scale as $\Theta(N \cdot N_{\text{fmax}})$ for DP5 and $\Theta(N \cdot N_{\text{rev}})$ for MP3. It can be seen that MP3 results in improvements on the order of 10 times less bandwidth for a long-term registration server. Short-term communication costs scale as $\Theta(N)$ for both protocols.

# 5. DISCUSSION

## 5.1 Scalability

A primary bottleneck in DP5 is its lack of scaling with large number of users, specifically for long-term epochs. MP3 solves just that. The complexity for bandwidth usage of all operations are summarized in Table 2.

The bulk of the cost in running a service such as MP3 or DP5 comes from the bandwidth usage of the given protocol. The bandwidth costs of MP3 are at most as expensive as DP5, and often cheaper. $N_{\text{rev}}$ is trivially always less than or equal to $N_{\text{fmax}}$, and in reality, with a 24-hour long-term epoch, setting $N_{\text{rev}} = 1$ or another small constant is very reasonable[6] making MP3 significantly cheaper during long-term epochs, and thus addressing the scalability issue that arose in DP5.

[6]Social networks such as Twitter disallow bulk unfollowing [25], making our argument about setting $N_{\text{rev}}$ to a small/constant value even stronger.

## 5.2 Explicit Buddy Revocations

Unlike DP5, MP3's revocation mechanism explicitly informs the revoked buddy that they have been revoked. To argue that DP5's revocation mechanism is implicitly finger printable, we analyzed the offline time of users on a popular public Internet Relay Chat (IRC) channel. Although IRC is not exactly a presence mechanism, it offers a good idea of users' online/offline statuses as a function of time.

We analyzed 6 months of IRC data[7] on the `#haskell` channel on the Freenode IRC network from January 1st through June 30th, 2016. Since our focus is presence, we recorded when users join and quit. Often, a user would temporarily disconnect without realizing, reconnect with a new nickname, and eventually change their nickname back to the original. We took this period of time as a single user being online for the entire duration and applied this correction in our analysis. We also only gave each user a single vote in the presence model as to not skew the result based on several users joining and quitting relatively often.

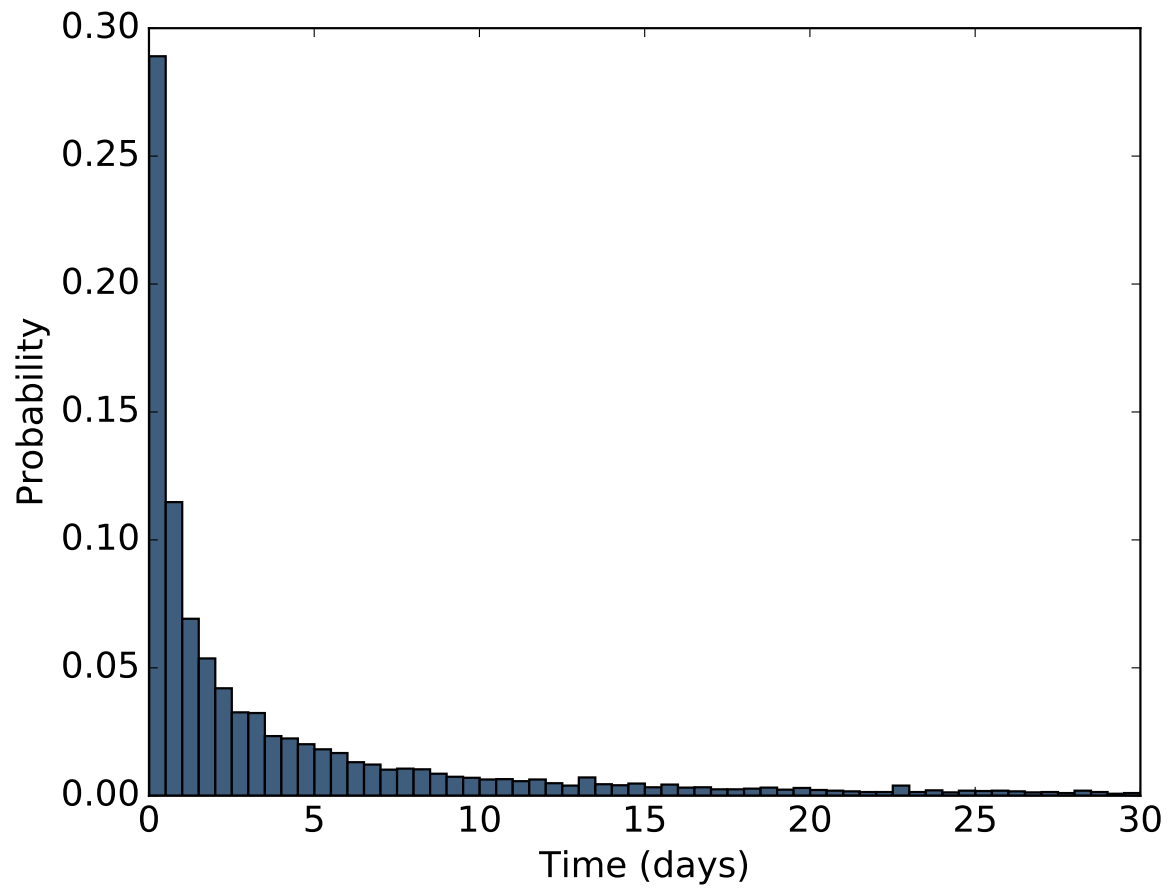

**Figure 5: PMF of user offline time analyzed over 6 months on the `#haskell` channel on the Freenode IRC network. Sample mean is 8.05 days.**

Figure 5 plots the probability mass function of user offline time in days. By taking the upper bound of the confidence interval for 99.9% confidence, we can say that the maximum offline time for any given user is about 8.74 days. In the context of DP5, this means that if you haven't seen one of your buddies come online after 9 long-term epochs, it is highly probable that you have been revoked.

[7]http://tunes.org/~nef/logs/haskell/

**Table 2: The complexities for bandwidth usage comparing MP3 and DP5 as a function of $N$, $N_{\mathbf{fmax}}$, and $N_{\mathbf{rev}}$.**

| | short-term | | | | long-term | | | |
|---|---|---|---|---|---|---|---|---|
| | Client | | Server | | Client | | Server | |
| | registration | lookup | registration | lookup | registration | lookup | registration | lookup |
| MP3 | $\Theta(1)$ | $\Theta(\sqrt{N})$ | $\Theta(N)$ | $\Theta(N\sqrt{N})$ | $\Theta(N_{\text{rev}})$ | $\Theta(\sqrt{N \cdot N_{\text{rev}}})$ | $\Theta(N \cdot N_{\text{rev}})$ | $\Theta(N\sqrt{N \cdot N_{\text{rev}}})$ |
| DP5 | $\Theta(1)$ | $\Theta(\sqrt{N})$ | $\Theta(N)$ | $\Theta(N\sqrt{N})$ | $\Theta(N_{\text{fmax}})$ | $\Theta(\sqrt{N \cdot N_{\text{fmax}}})$ | $\Theta(N \cdot N_{\text{fmax}})$ | $\Theta(N\sqrt{N \cdot N_{\text{fmax}}})$ |

A more sophisticated analysis could be done in the context of one's own buddies using the knowledge of their behavior and habits in the context of online presence. For example, if you have a buddy that you know only comes on Wednesday mornings and Saturday evenings, you would sense an anomaly with that particular friend had they not come online after a certain number of weeks. Using this kind of logic, a presence model of your own set of friends could be derived based on their known habits.

It is also often the case that friend groups overlap, that is, if $A$ is friends with $B$ and $B$ is friends with $C$, it is highly likely that $A$ is friends with $C$. This means that if $A$ has any concern that $B$ has revoked them, they can simply ask $C$ if $B$ has been online recently.

These arguments show that the indistinguishability of offline status and revocation in DP5 is unnecessary and the explicit revocations of MP3 are not so much a drawback.

## 5.3 Malicious Parties

Some conventional approaches to ensure availability against malicious parties cannot be applied directly to privacy-preserving protocols, as they can leak information. This causes several challenges: keeping the databases small, ensuring the registration server stores all uploaded presence records, ensuring the lookup servers store all presence records and do not modify them.

A malicious client could upload many presence records during a given epoch, causing a denial of service (DoS) for all other clients. If a malicious client were using an anonymous channel, authentication would compromise the anonymity of that client, defeating the purpose of MP3. To eliminate this, $k$-times anonymous authentication schemes have been proposed [26–29]. In these schemes, users are guaranteed anonymity up to $k$ times; that is, if a user authenticates $k+1$ times, the identity of the user can be computed. Another solution is using $n$-time anonymous authentication [30]. Such private rate-limiting schemes can be used to limit the number of times a client registers during a given epoch without losing anonymity.

In the case that the registration server is dishonest and drops records, a user could "friend themself" to ensure that their presence records are being stored, by looking themself up during every epoch. Note that all presence records are indistinguishable, so the registration server can not target specific records for dropping.

Lastly, in the case that the lookup servers are dishonest and modify the database, Devet et al. propose a robust PIR scheme [9] that allows detection of malicious servers. This detection requires at least $t+2$ honest servers, where $t$ is the maximum number of servers allowed to collude to be able to determine the data in a query. This robust PIR scheme is implemented in MP3.

## 5.4 Cost Improvements

As stated in Section 5.1, the main component contributing to the cost of a service such as MP3 or DP5 comes from the bandwidth usage, with the most expensive part being a long-term lookup server. For $N = 10000$, $N_{\text{fmax}} = 100$, and $N_{\text{rev}} = 1$, and assuming the arguments in Section 3.5.3, the cost of running an MP3 long-term lookup server for one month (30 days) is on the order of 400 GB of bandwidth and the cost of running a DP5 long-term lookup server for one month is on the order of 900 GB of bandwidth. This difference grows faster as $N$ increases due to the $\Theta(N\sqrt{N})$ scaling (holding other parameters fixed) of a lookup server's bandwidth requirements. So MP3 provides a great improvement to DP5's long-term lookup.

MP3 has also greatly reduced the registration cost as compared with DP5 due to registration messages growing with $\Theta(N_{\text{rev}})$ for MP3 and $\Theta(N_{\text{fmax}})$ for DP5. This gives an improvement on the order of 10 times that of DP5 for 10000 users. This has less impact on the total cost of the service, though, as registration costs for both protocols are on the order of megabytes, while lookup costs are on the order of gigabytes.

It should also be noted that $N_{\text{fmax}}$ is on the order of $N$, while $N_{\text{rev}}$ is on the order of a constant. This means that the bandwidth costs of a long-term DP5 lookup server grow approximately as $\Theta(N^2)$ while the bandwidth costs of a long-term MP3 lookup server grow approximately as $\Theta(N\sqrt{N})$.

# 6. CONCLUSION

We propose MP3, an efficient private presence protocol that leaks no information about the graph structure of the social network. We improved the existing DP5 private presence protocol while preserving the same level of privacy by decreasing the overall bandwidth usage required of the service.

We improved the most expensive part of DP5, the long-term registration and lookup by leveraging a dynamic broadcast encryption scheme to keep long-term databases small, allowing for cheap queries in both the long-term and short-term. In reference to DP5, MP3 this requires on the order 10 times less bandwidth during long-term registration, and on the order 2 times less bandwidth during long-term lookups for 10000 users, and this difference only grows larger as the number of users grows as the bandwidth costs scale with $\Theta(N)$ for registration and $\Theta(N\sqrt{N})$ for lookup.

## Acknowledgments

This research was partially supported by the NSF under grant 1314637 and by the University of Minnesota Undergraduate Research Opportunities Program.

## 7. REFERENCES

[1] D. Rushe. Lavabit founder refused FBI order to hand over email encryption keys. The Guardian, October 2013.

[2] Apple. Approach to privacy. http://www.apple.com/privacy/approach-to-privacy/.

[3] Open Whisper Systems. https://whispersystems.org/.

[4] M. Marlinspike. Facebook messenger deploys signal protocol for end to end encryption. https://whispersystems.org/blog/facebook-messenger.

[5] N. Borisov, G. Danezis, and I. Goldberg. DP5: A private presence service. *Proceedings on Privacy Enhancing Technologies*, 2015(2):4–24, 2015.

[6] B. Chor, O. Goldreich, E. Kushilevitz, and M. Sudan. Private information retrieval. In *IEEE Symposium on Foundations of Computer Science*, 1995.

[7] C. Delerablée, P. Paillier, and D. Pointcheval. Fully collusion secure dynamic broadcast encryption with constant-size ciphertexts or decryption keys. In *Pairing-Based Cryptography–Pairing 2007*, pages 39–59. Springer, 2007.

[8] W. Diffie and M. Hellman. New directions in cryptography. *IEEE transactions on Information Theory*, 22(6):644–654, 1976.

[9] C. Devet, I. Goldberg, and N. Heninger. Optimally robust private information retrieval. In *Presented as part of the 21st USENIX Security Symposium (USENIX Security 12)*, pages 269–283, 2012.

[10] D. I. Wolinsky, H. Corrigan-Gibbs, B. Ford, and A. Johnson. Dissent in numbers: Making strong anonymity scale. In *Presented as part of the 10th USENIX Symposium on Operating Systems Design and Implementation (OSDI 12)*, pages 179–182, 2012.

[11] H. Corrigan-Gibbs, D. Boneh, and D. Mazières. Riposte: An anonymous messaging system handling millions of users. In *2015 IEEE Symposium on Security and Privacy*, pages 321–338. IEEE, 2015.

[12] N. Koblitz and A. Menezes. *Pairing-based cryptography at high security levels*. Springer, 2005.

[13] D. Boneh, X. Boyen, and H. Shacham. Short group signatures. In *Annual International Cryptology Conference*, pages 41–55. Springer, 2004.

[14] J. Lopez and R. Dahab. An overview of elliptic curve cryptography. 2000.

[15] D. J. Bernstein, N. Duif, T. Lange, P. Schwabe, and B.-Y. Yang. High-speed high-security signatures. *Journal of Cryptographic Engineering*, 2(2):77–89, 2012.

[16] D. Boneh, B. Lynn, and H. Shacham. Short signatures from the weil pairing. In *International Conference on the Theory and Application of Cryptology and Information Security*, pages 514–532. Springer, 2001.

[17] A. Fiat and M. Naor. Broadcast encryption. In *Advances in Cryptology—CRYPTO '93*, pages 480–491. Springer, 1993.

[18] I. Goldberg. Improving the robustness of private information retrieval. In *2007 IEEE Symposium on Security and Privacy*, pages 131–148. IEEE, 2007.

[19] B. Chor, N. Gilboa, and M. Naor. *Private information retrieval by keywords*. Citeseer, 1997.

[20] J. Daemen and V. Rijmen. *The design of Rijndael: AES-the advanced encryption standard*. Springer Science & Business Media, 2013.

[21] D. Eastlake 3rd and P. Jones. US secure hash algorithm 1 (SHA1). Technical report, 2001.

[22] D. F. Aranha and C. P. L. Gouvêa. RELIC is an Efficient LIbrary for Cryptography. https://github.com/relic-toolkit/relic.

[23] I. Goldberg, C. Devet, W. Lueks, A. Yang, P. Hendry, and R. Henry. Percy++ project on sourceforge. http://percy.sourceforge.net, 2014.

[24] D. McGrew and J. Viega. The security and performance of the galois/counter mode (GCM) of operation. *Progress in Cryptology-INDOCRYPT 2004*, pages 377–413, 2005.

[25] Suspension - what's the daily/hourly unfollow limit for each user? what is the aggressive behaviour? https://twittercommunity.com/t/suspension-whats-the-daily-hourly-unfollow-limit-for-each-user-what-is-the-aggressive-behaviour/13971.

[26] J. K. Liu, V. K. Wei, and D. S. Wong. Linkable spontaneous anonymous group signature for ad hoc groups. In *Australasian Conference on Information Security and Privacy*, pages 325–335. Springer, 2004.

[27] P. P. Tsang and V. K. Wei. Short linkable ring signatures for e-voting, e-cash and attestation. In *International Conference on Information Security Practice and Experience*, pages 48–60. Springer, 2005.

[28] P. P. Tsang, V. K. Wei, T. K. Chan, M. H. Au, J. K. Liu, and D. S. Wong. Separable linkable threshold ring signatures. In *International Conference on Cryptology in India*, pages 384–398. Springer, 2004.

[29] P. P. Tsang, M. H. Au, A. Kapadia, and S. W. Smith. Blacklistable anonymous credentials: blocking misbehaving users without TTPs. In *Proceedings of the 14th ACM conference on Computer and communications security*, pages 72–81. ACM, 2007.

[30] J. Camenisch, S. Hohenberger, M. Kohlweiss, A. Lysyanskaya, and M. Meyerovich. How to win the clonewars: efficient periodic n-times anonymous authentication. In *Proceedings of the 13th ACM conference on Computer and communications security*, pages 201–210. ACM, 2006.